\documentclass[12pt,letterpaper]{article}
\usepackage{jheppub}
\usepackage{bbold}

\title{Complementarity Is Not Enough}

\author[a,b]{Raphael Bousso}
\affiliation[a]{Center for Theoretical Physics and Department of Physics,\\
 University of California, Berkeley, CA 94720, U.S.A.}
\affiliation[b]{Lawrence Berkeley National Laboratory, Berkeley, CA 94720,
  U.S.A.}

\abstract{%
The near-horizon field $B$ of an old black hole is maximally entangled with the early Hawking radiation $R$, by unitarity of the S-matrix.  But $B$ must be maximally entangled with the black hole interior $A$, by the equivalence principle.  Causal patch complementarity fails to reconcile these conflicting requirements. The system $B$ can be probed by a freely falling observer while there is still time to turn around and remain outside the black hole.  Therefore, the entangled state of the $BR$ system is dictated by unitarity even in the infalling patch.  If, by monogamy of entanglement, $B$ is not entangled with $A$, the horizon is replaced by a singularity or ``firewall''. 

To illustrate the radical nature of the ideas that are needed, I briefly discuss two approaches for avoiding a firewall: the identification of $A$ with a subsystem of $R$; and a combination of patch complementarity with the Horowitz-Maldacena final-state proposal.}
\begin{document}
\maketitle

\section{The Firewall Paradox}

Almheiri, Marolf, Polchinski and Sully~\cite{AMPS} have exhibited a profound paradox.  Consider a Schwarzschild black hole that formed from a pure state and which has partially evaporated, at the Schwarzschild time $t=0$.  Let $R$ be the early Hawking radiation, which was emitted while $t<0$.  Let $R'$ be the late Hawking radiation, which will be produced when $t>0$, by what remains of the black hole.  

By unitarity~\cite{SusTho93,StrVaf96,Mal97}, the whole system $RR'$ is a pure state in a Hilbert space of dimension $e^N$, where $N$ is given by the horizon entropy~\cite{Bek72,Haw75} of the newly formed black hole: $N={\cal A}_0/4$.  Now suppose that at the time $t=0$ of a distant observer, the black hole is older than the Page time, of order $r_S^3$, so that its area is less than half of its original size: ${\cal A} <{\cal A}_0$.  Then the early radiation $R$ is the larger subsystem, $\log\dim {\cal H}_ R>N/2$; and the late radiation is the smaller subsystem, $\log\dim {\cal H}_{R'}<N/2$.  In a typical pure state, the entanglement entropy of a smaller subsystem is nearly maximal~\cite{Pag93}:
\begin{equation}
S_{RR'}=0;~~S_{R'}= \log\dim {\cal H}_{R'}-\epsilon~,
\label{eq-srrp}
\end{equation}
where $S_{R'}=-{\rm Tr}\rho_{R'} \log \rho_{R'} $ is the von Neumann entropy of $\rho_{R'}$, the density matrix obtained from the global pure state by a partial trace over ${\cal H}_R$; and $\epsilon>0$ is exponentially small in the difference between the subsystem sizes, $\log\dim {\cal H}_ R-\log\dim {\cal H}_{R'}$.  

At $t=0$, the Hilbert space factor $R'$ is just the remaining black hole, which has not yet dissolved into the late Hawking radiation.  Thus, an old black hole is (nearly) maximally entangled with the radiation it has already emitted.  

Let us now review what complementarity states about the description of a black hole (young or old).  Complementarity distinguishes the viewpoint of an observer who remains far from the black hole, Bob, from that of an infalling observer, Alice.  These viewpoints have to be consistent as long as they can be operationally compared.

From Bob's viewpoint, a black hole can be thought of as an object consisting of two subsystems which constantly interact~\cite{SusTho93}:
\begin{equation}
R'=BH~.
\label{eq-rpbh}
\end{equation}
$H$ is called the stretched horizon, a membrane of Planckian thickness above the true event horizon.  $B$ is the near horizon zone (``the zone''), a shell of proper width of order $r_S$ just outside the membrane.  Operationally, the difference between $H$ and $B$ is that $B$ can be probed by Bob without experiencing accelerations greater than the Planck scale, while $H$ cannot.

Let us take the infalling observer, Alice, to be small compared to $r_S$.  She enters the near-horizon zone (of order $r_S$ from the horizon) in free fall at the time $t=0$. Near the horizon, on time and distance scales much less than $r_S$, Alice can approximate the metric as that of Minkowski space.  Assuming that no matter is falling in along with her, Alice should perceive the vacuum of Minkowski space on these scales.  

Minkowski space can be divided into a left and right Rindler wedge.  The vacuum state is maximally entangled between fields with support in the two wedges~\cite{Unr76}.  Locally, the black hole horizon can be identified with a Rindler horizon, and $B$ can be identified with the right Rindler wedge.  Therefore, Alice must find any modes that are localized outside the horizon to better than $r_S$ to be maximally entangled with similarly localized modes $A$ inside the horizon.   ($B$ denotes both the near horizon region and the quantum fields it contains.  Similarly, $A$ will denote both the black hole interior and the fields with support in it.)

In summary, the equivalence principle\footnote{See Sec.~\ref{sec-fire} for a more detailed discussion of the sense in which firewalls can be regarded as a violation of the equivalence principle.} applied to the freely falling observer Alice requires that the system $AB$ is approximately in a maximally entangled pure state, the local Minkowski vacuum:
\begin{equation}
S_{AB}=0;~~~S_B\approx \log\dim {\cal H}_B~.
\label{eq-sab}
\end{equation}

By Eq.~(\ref{eq-rpbh}), $B\subset BH$ is a subsystem of the black hole at the time $t=0$, in the description of Bob.  By unitarity, it follows that $B$ is a subsystem of the late Hawking radiation $R'$ that the black hole evolves into.  By Eq.~(\ref{eq-srrp}), there exists a subsystem $R_{BH} \subset R$ of the early Hawking radiation that purifies $R'=BH$.  It follows that there exists a (smaller) subsystem $R_B\subset R_{BH}\subset R$ that purifies $B\subset BH$:
\begin{equation}
S_{R_B B}=0~.
\label{eq-rbb}
\end{equation}
This conflicts with Eq.~(\ref{eq-sab}).  A system (in this case $B$) cannot be maximally entangled with two distinct other systems.  Formally, this violates the subadditivity of the entropy~\cite{AMPS}.  

This is the paradox.  If we insist on unitarity, Eq.~(\ref{eq-srrp}), then we are forced to abandon the equivalence principle, Eq.~(\ref{eq-sab}).  The field theory degrees of freedom just inside and outside the black hole cannot be mutually entangled or even correlated.  This implies that an infalling observer experiences arbitrarily high energy quanta near the horizon: a firewall.

\section{Why Complementarity is Not Enough}
\label{sec-notenough}

Naively~\cite{Bou12c}, the firewall paradox can be resolved by fully exploiting the freedom offered by complementarity: Alice and Bob can have different theories for predicting their observations.  Each theory must be consistent with quantum mechanics, and with semi-classical gravity in its regime of validity.  But the theories need only agree on observations that Alice and Bob can communicate without violating causality or leaving the regime of semi-classical gravity.

In a general spacetime, consider all inextendible worldlines.  They will fall into equivalence classes defined by the intersection of the past and the future of a given worldline.  Such a set will be called a causal patch. Causal patches are the fundamental objects of complementarity in arbitrary spacetimes~\cite{Bou00a}, and the covariant entropy bound~\cite{CEB1} naturally applies to their boundaries.  (See also Refs.~\cite{BanFis01a,BanFis01b,BanFis04b,Ban10a,Ban11}.)  Every causal patch must have a consistent description, and if transplanckian accelerations are needed to explore certain regions of the patch, then the semiclassical regime is restricted to the interior of a stretched horizon~\cite{SusTho93}.   

For example, what Alice's theory predicts for her observations at or behind the stretched horizon cannot be communicated to Bob, because this region is outside of Bob's causal patch.  Another way of saying this is that at such late times, she no longer has a choice whether she wants to end up as an outside or inside observer. The theory describing her observations then need not be consistent with Bob's theory, and the combination of both theories into a global picture may yield a contradiction.

A classic example is the resolution of the quantum xeroxing problem~\cite{Pre92,SusTho93b,HayPre07}.  At sufficiently late times in the theory of a collapsing star (Alice), the pure state of the star is located well inside the black hole (but still far from the singularity).  At late times in Bob's theory the same pure state is located in the Hawking radiation.  The naive combination of both descriptions into a single global geometry leads to a contradiction: the quantum state of the star has been cloned, in violation of the linearity of quantum mechanics.  But no observer can see both copies.  Complementarity resolves the cloning paradox.

A similar type of resolution can be envisaged for the firewall paradox~\cite{Bou12c}.  Both Alice and Bob have equal access to the early Hawking radiation $R$ and must agree on its state.  When Bob examines the late Hawking radiation, $R'$, he will find that it is purified by a subsystem of $R$, as demanded by unitarity.  He does not have access to the black hole interior $A$.  Therefore he cannot establish a contradiction by verifying that a subsystem of $R'$ is also purified by a different system, $A$. In short, Bob can verify Eq.~(\ref{eq-srrp}), so in particular he can verify Eq.~(\ref{eq-rbb}).  But he cannot verify its contradiction, Eq.~(\ref{eq-sab}).

Alice, on the other hand, jumps into the black hole and thus cannot measure the late Hawking radiation, $R'$.  Therefore, she cannot verify Eq.~(\ref{eq-srrp}) directly, and thus establish a conflict between it and Eq.~(\ref{eq-sab}).  She can experience the vacuum at the horizon, but by then it is too late to tell Bob, or (equivalently) to fire her rockets and become like Bob, a distant observer at late times.

In order for this resolution to be valid, it must pass a consistency check articulated by Harlow~\cite{Har12}: it must be impossible for Alice to measure $B$ {\em before\/} she reaches the stretched horizon.  Otherwise, Alice could measure the relevant subset of the late Hawking radiation on her way to the horizon, in its incarnation as the near-horizon modes $B\subset BH=R'$.  At this point she could still decide to fire her rockets and stay outside, so her theory must agree with Bob's theory.  It must predict that $B$ is maximally entangled with the early radiation, Eq.~(\ref{eq-rbb}).  But then her theory cannot also predict that $B$ is maximally entangled with the modes inside the horizon, Eq.~(\ref{eq-sab}).  If Alice can measure $B$ before crossing the horizon, then complementarity is not enough to evade the firewall argument.

At first, it might appear that such measurements are difficult for Alice.  If she wants to remain in the near horizon zone for a long time, she must accelerate outward, which might pollute the setup due to emissions from her detector.  If she wants to measure $B$ while in free fall, then the modes in question are comparable in wavelength to the time she has left before crossing the horizon~\cite{Har12,MatThu12}.  This limits the precision with which $B$ can be measured. 

However, in the limit of a large black hole ($r_S\to\infty$), none of these complications appear to rise to the level of an in-principle obstruction\footnote{I thank D.~Marolf and J.~Polchinski for insisting on this point in a number of communications.  I am grateful to B.~Freivogel, D.~Harlow, S.~Leichenauer, D.~Stanford, and L.~Susskind  for extensive discussions of Alice's ability to examine $B$ before reaching the horizon.} to measuring arbitrarily many q-bits in $B$ and gaining arbitrarily good statistics establishing their correlation with $R_B$.  In particular, the validity of the firewall argument does not rest on the ability to measure any particular near-horizon mode with arbitrarily high accuracy; some finite fidelity is sufficient.

It is possible that a fundamental obstruction to the measurement of near-horizon modes $B$ prior to horizon crossing arises from some constraint that has been overlooked.  But for now, it is reasonable to conclude that the consistency check {\em fails}.  Complementarity is not enough.

\section{Three Choices}
\label{sec-discussion}

Does this mean that there are firewalls?  Not necessarily.  But unless a further, hidden assumption can be identified, the following three postulates are not mutually consistent~\cite{AMPS}:
\begin{itemize} 
\item The formation and evaporation of a black hole is a unitary process.
\item An infalling observer sees nothing special at the horizon.
\item Effective quantum field theory is valid outside the stretched horizon.
\end{itemize} 
One of the postulates has to be given up.  I will discuss each possibility in turn.

The following discussion will be colored by my own theoretical prejudices and expectations.  There is currently a large spectrum of views on the subject of firewalls%
~\cite{Sus12b,Sus12c,NomVar12,MatThu12,ChoPuh12,BenPuh12,GivItz12,BanFis12,Ori12,Bru12}; the latest version of Ref.~\cite{AMPS} responds to several of these works.

\subsection{Information Loss for the Outside Observer}
\label{sec-infoloss}

Proponents of information loss have every right to feel vindicated by the firewall paradox.  Yet, I find this possibility implausible.  There is overwhelming evidence that the full nonperturbative quantum gravity theory for certain asymptotically Anti-de Sitter spacetimes is a unitary conformal field theory~\cite{Mal97}.  This implies that black holes in Anti-de Sitter space (which include black holes in nearly flat regions) evolve unitarily~\cite{Mal97}.  Information might remain stuck in small remnants with arbitrarily large entropy, but such objects would seem to lead to instabilities due to their large phase space.  They are inconsistent with entropy bounds~\cite{Tho93,Sus95,CEB1} in general, and with the UV/IR relation of AdS/CFT~\cite{SusWit98} in particular.   

There is a more general argument for unitarity, which I find compelling.  It is based on the principle that laws of physics must be operationally meaningful.  Bekenstein~\cite{Bek72} argued that black holes must carry entropy, because otherwise the second law of thermodynamics would be ``transcended''.  That is, the second law would be operationally meaningless, since the matter inside a black hole cannot be accessed and its entropy is effectively lost.  This was a daring argument to make---the obvious but wrong answer would have been to concede that matter and its entropy are simply lost into the black hole---but it carried the day~\cite{Haw75}.  

Entropy and information are two sides of the same coin, and if entropy cannot be lost down a black hole, then it would be bizarre if information could be lost in this way.  Indeed, Bekenstein's argument is just as compelling if one replaces ``entropy'' (i.e., lack of information about the fine-grained quantum state) by ``information'', and ``second law'' by ``unitarity''.  If information could be lost into a black hole, then the law of unitarity would be transcended.  The law would be operationally meaningless, because no-one could verify that it still holds after matter enters a black hole. I expect that unitarity (for the outside observer) will be upheld as an operationally meaningful law of Nature. 

\subsection{Firewalls} 
\label{sec-fire}

Why not give up the postulate of harmless horizon crossing, and explore the possibility that firewalls really exist?  It would be interesting to beyond an argument-by-contradiction, and to gain a more direct understanding of their origin and dynamics.  It is unclear, for example, whether firewalls would first form at the scrambling time~\cite{AMPS}, of order $r_S\log r_S$, or at the much later Page time~\cite{Sus12b}, when half the black hole's area has evaporated.

An extension of the firewall argument~\cite{AMPS} to young black holes suggests that a mild firewall would already form at the scrambling time.  The number of excited modes seen by an infalling observer at this time would be set by the logarithm of the dimension of the Hilbert space of the collapsed matter system, which is typically much smaller than the black hole area.  Thus, the putative initial firewall may be unobservable in practice.\footnote{I thank D.~Marolf for discussions of this point.}  As the black hole evaporates, its initial pure state evolves into an incoherent superposition of more typical black hole states, purified by the emitted radiation. The firewall would grow continuously during this period and reach maximum size at the Page time.  I obtain this conclusion by assuming that the stretched horizon dissolves into the inside modes $A$ when an observer falls through it, and that its state mimics the vacuum to the extent consistent with the requirement that it also store the information about the infalling matter system.  I also assume, for definiteness, that the stretched horizon and the near horizon region (the ``zone'') are each responsible for half of the Bekenstein-Hawking entropy.\footnote{A similar conclusion was recently obtained by L.~Susskind (private communication and Ref.~\cite{Sus12c}).}

But there are good reasons to abhor firewalls:
\begin{itemize} 

\item Firewalls signal a failure of the equivalence principle.  Outside the horizon of an old black hole, the metric is given by the Kerr-Newman solution, and the curvature radius can be arbitrarily large.  By the equivalence principle, an inertial observer much larger than the Planck size and much smaller than the curvature radius should be able to approximate the local spacetime near the horizon as Minkowski space.  Yet, spacetime would effectively end at the horizon.  Only a null shockwave of Planck density could modify the classical solution so dramatically.  So far, the only origin for this type of matter source that has been proposed is the firewall argument itself: effectively, the vacuum disintegrates over the Page time through the loss of entanglement.  An inertial observer would have no knowledge of the presence and location of the firewall unless she has carefully monitored spacetime on the Page timescale and the horizon distance scale, both of which can be arbitrarily large.  This amounts to the breakdown of the equivalence principle for all practical purposes.  

\item Firewalls do not appear to change the conclusion that a global description of the universe fails in causally nontrivial spacetimes.  This situation is extremely uneconomical.  We are left with two radical and apparently independent modifications of physics in order to solve the single problem of reconciling unitarity of the Hawking radiation with the no-cloning theorem.  Certain attempts to verify quantum xeroxing are now thwarted by {\em two} independent obstructions.  The scrambling time already delays Hayden-Preskill mirroring of information just enough to avoid the xeroxing paradox~\cite{HayPre07}; but an observer who jumps in before this time fails to see xeroxing for two reasons: no outside copy has been generated in the Hawking radiation, and the inside copy is inaccessible due to the firewall.  This is overkill.  Yet, it appears that we cannot abandon complementarity.  In the global picture, there are still two copies of a single quantum state.  The firewall forms no earlier than the scrambling time, so it is possible to accompany a collapsing star and observe its quantum state inside the black hole.  Unitarity demands that the same state is present in the Hawking radiation.  Both copies are in the future of the uncollapsed star, and neither is in the future of the other. Thus the restriction to a single causal patch remains important in order to avoid xeroxing.  (Another argument for the continued importance of complementarity was sketched by Susskind~\cite{Sus12b}.)

\end{itemize} 
If there are no firewalls,\footnote{Another concern about firewalls, not listed above, was emphasized to me by J.~Maldacena: the absence of an interior spacetime invalidates Hawking's calculation~\cite{Haw75}.  The semiclassical evolution of the black hole after the formation of the firewall is thus out of control~\cite{AMPS}, similar to the appearance of a naked singularity.  For all we know, according to the firewall picture, a large black hole might explode.  However, that the Second Law may prevent such explosions, if it and the Bekenstein formula remain valid at firewalls~\cite{Bro12} (D.~Marolf, private communication).} and if unitarity is preserved for the outside observer, then there is only one remaining option.

\subsection{Breakdown of Effective Quantum Field Theory} 

Based on the above assessments, it seems reasonable to explore the consequences of rejecting both information loss and firewalls.  Complementarity is not sufficient to render these assumptions consistent with unitarity, because Alice's theory alone is not consistent as it stands.  Before she enters the black hole, Eq.~(\ref{eq-sab}) must hold if she is to avoid a firewall while entering. But the incompatible Eq.~(\ref{eq-rbb}) must also hold since at that same time she still has the option of remaining outside the black hole and verifing unitarity.  Thus, effective quantum field theory must break down outside the stretched horizon, at least for an infalling observer.  

In the remainder of this note, I will briefly examine two directions that may be explored.  Both are speculative, but they illustrate my expectation that if firewalls can be avoided, dramatic new physics is needed.\footnote{Certain limited modifications of effective field theory~\cite{Gid11,Gid12,GidShi12} were argued to be insufficient in Ref.~\cite{AMPS}.}  In particular, my hope is that we will finally learn something about the fundamental description of the infalling observer, and thus, perhaps, about questions such as the validity of quantum mechanics in cosmology, and the effective loss of information near a singularity.

\subsubsection{Identification of Hilbert Spaces}

One way to make the two equations consistent is to identify the two apparently distinct systems that purify the near horizon modes $B$:\footnote{I thank Douglas Stanford for discussions that led me to explore this possibility.}
\begin{equation}
{\cal H}_A={\cal H}_{R_B}~.
\label{eq-arb}
\end{equation}
The interior of the black hole, $A$, is fundamentally the same object as a certain Hilbert space factor of the early Hawking radiation, $R_B\subset R$.

This has a familiar ring to it---black hole complementarity has often been phrased in terms of an identification of Hilbert spaces inside and outside the black hole.  But in the context of the firewall argument, the above assertion is more specific and thus nontrivial.  There are three consistency requirements that it must survive, and an additional difficulty.

First, $A$ and $R_B$ should not be simultaneously accessible to Alice as physically distinct systems.  Generically, they are not: by entropy bounds, it is impossible to carry all of the early radiation into the near horizon region.  But what if Alice extracts a small subsystem $R_b\subset R_B$ from the Hawking radiation, entangled with some particular modes $b\subset B$, and transports it to the black hole?  Extraction of $R_b$ is extremely difficult, since $R_b$ is highly scrambled in $R$, and the presence of firewalls for highly nongeneric observers may be acceptable.  It is possible, after all, to create arbitrary distant excitations (with some nonzero probability) by making appropriate measurements in a region of Minkowski space.  The crucial question, which I will not address here, is whether the coherent transport of $R_b$ into the black hole can lead to contradictions akin to quantum xeroxing, if $A$ and $R_B$ are identified.

Secondly, the quantum information in $R_B$ must somehow get into the black hole and become $A$, without violating causality, and without violating the previous requirement.  I am assuming that Alice is generic (i.e., ignorant): she does not carry any subsystem of $R_B$ with her into the black hole.  Then the only causal channel for $R_B$ is along the boundary of Alice's patch.  In some sense, Alice's patch boundary would have to store $R_B$.  This behavior would be quite remarkable, since $R_B$ passes out of Alice's patch through a component of her boundary that is not a Killing horizon, whose area does not respond like a black hole by increasing; it is rapidly decreasing.  If this boundary is nevertheless assigned a Hilbert space, presumably of dimension comparable to the exponential of its area, the additional problem arises that at the time when $R_B$ becomes $A$, the boundary area is not much larger than the number of q-bits in $R_B$.   Therefore, Alice's patch boundary would have to store $R_B$ {\em selectively\/}, at the expense of the arbitrarily large amount of other information that passes through it at early times.\footnote{It is tempting to propose that the boundary state always purifies the bulk, including the entanglement entropy of the vacuum.  This would generalize the unitary behavior of Killing horizons to non-Killing horizons.  But such behavior appears to conflict with standard quantum mechanics. I thank D.~Marolf and J.~Polchinski for discussions of this point.}

The third condition is that the patch boundary finally releases $R_B$ as bulk modes with support inside the black hole.  This behavior is at least qualitatively comparable to properties that are already ascribed to black hole horizons in the standard picture of complementarity.  In both cases, the details of the dynamics are complicated but would have to be dictated by the operation of basic principles (in this case, unitarity and free fall).  Instead of waiting for a long time for the stretched horizon to dissolve into Hawking radiation, Alice would cause it to dissolve rapidly into $A$ when she crosses it.

If all of the above requirements can be met, it may be consistent to identify the Hilbert spaces of $A$ and $R_B$.  However, there would still be at least a small firewall, if the mapping between states was one-to-one.  The state in the Hilbert space ${\cal H}_B\otimes {\cal H}_{R_B}$ is an arbitrary pure state $|\psi \rangle$.  But the infalling vacuum is not just any pure, maximally entangled state in ${\cal H}_A\otimes {\cal H}_B$; it is the particular pure state $|0\rangle$.  There are two possibilities: Either there exists a small firewall at all times, whose size is set by the dimension of the matter Hilbert space that formed the black hole and which may in practice be unobservable;  or the mapping between the states in the $A$ and $R_B$ Hilbert spaces is many-to-one.

\subsubsection{Final State Quantum Mechanics}
\label{sec-finalstate}

Horowitz and Maldacena (HM)~\cite{HorMal03} proposed a way to reconcile unitarity with the assumption~\cite{Haw75} that the state in the vicinity of the horizon (as seen by an inertial observer away from infalling matter) is the vacuum.  The HM proposal is based on a generalization of ordinary quantum mechanics that allows for the specification of a final state in addition to an initial state~\cite{AhaBer64,GelHar91}; ordinary quantum mechanics is recovered as the special case where the final state is proportional to the unit density matrix.

The final state in the HM proposal is specified only for the Hilbert space factors of the infalling matter system, $M$, and of the quantum fields supported inside the black hole, and it is taken to be a particular maximally entangled state in the product space ${\cal H}_M\otimes {\cal H}_{\rm in}$.  In the Hilbert space factor corresponding to the Hawking radiation, ${\cal H}_{\rm out}$, no final state is specified.  Thus\footnote{The notation and conventions follow Gottesman and Preskill~\cite{GotPre03}, where more details can be found.}
\begin{equation}
\rho_f= |\Psi \rangle_{M\otimes{\rm in}} \mbox{~}_{M\otimes{\rm in}}\langle \Psi|\otimes \mathbb{1}_{\rm out}~,
\end{equation}
where 
\begin{equation}
\mbox{}_{M\otimes{\rm in}}\langle \Psi|= \mbox{}_{M\otimes{\rm in}}\langle \Phi|  (S_M\otimes \mathbb{1}_{\rm in})~,
\label{eq-final}
\end{equation}
and
\begin{equation}
\mbox{}_{M\otimes{\rm in}}\langle \Phi| = N^{-1/2} \sum_{i=1}^N \mbox{}_M\langle i|\otimes \mbox{}_{\rm in} \langle i|~.
\end{equation}
$S_M$ is a unitary operator that effectively becomes the S-matrix to the outside observer.  $\{|i \rangle_X\}$ is an orthonormal basis in the Hilbert space ${\cal H}_X$; and all three Hilbert spaces are taken to have equal dimension $N=e^{{\cal A}_0/4}$, where ${\cal A}_0$ is the initial area of the black hole.

The initial state of the matter system is an arbitrary pure state $|\psi\rangle_M$.  The initial state for the in and out sectors is maximally entangled so as to form the infalling vacuum:
\begin{equation}
|\Phi \rangle_{{\rm in}\otimes {\rm out}} = N^{-1/2} \sum_{i=1}^N |i \rangle _{\rm in}\otimes |i \rangle _{\rm out} 
\end{equation}
The effect of the final state can be thought of as the quantum teleportation~\cite{BenBra92} of the pure matter state $|\psi\rangle_M$ to the Hawking radiation.  Unlike in quantum teleportation, the outcome of the Bell measurement on ${\cal H}_M\otimes {\cal H}_{\rm in}$ is determined by the theory to be the particular Bell state state such no further unitary operation on ${\cal H}_{\rm out}$ is needed in order to obtain the correct out state $S_M |\psi\rangle_M$. Therefore, no classical information need be sent to the asymptotic observer. 

From the viewpoint of complementarity, one might have questioned whether the problem addressed by HM is really present.  An outside observer cannot probe the interior of the black hole or the behavior of the horizon under free-fall.  To him, the Hilbert space ${\cal H}_{\rm in}$ is operationally meaningless.  Complementarity instructs us that the unitary S-matrix is the result of (quantum-mechanically) conventional evolution of the matter system to a stretched horizon and then to a Hawking cloud.

However, the firewall argument demonstrates that a conflict with ordinary quantum mechanics does arise for an observer Alice who falls into an old black hole, with area ${\cal A} \leq {\cal A}_0/2$: the monogamy of entanglement is violated.  Harmless free fall requires that the near-horizon modes ${\cal H}_B\subset H_{\rm out}$ be maximally entangled with degrees of freedom ${\cal H}_A\subset {\cal H}_{\rm in}$ inside the black hole.  But unitarity requires that the same system ${\cal H}_B$ be maximally entangled with its complement in $H_{\rm out}$, the early Hawking radiation ${\cal H}_R$.
\begin{figure}
\begin{center}
\includegraphics[scale = .43]{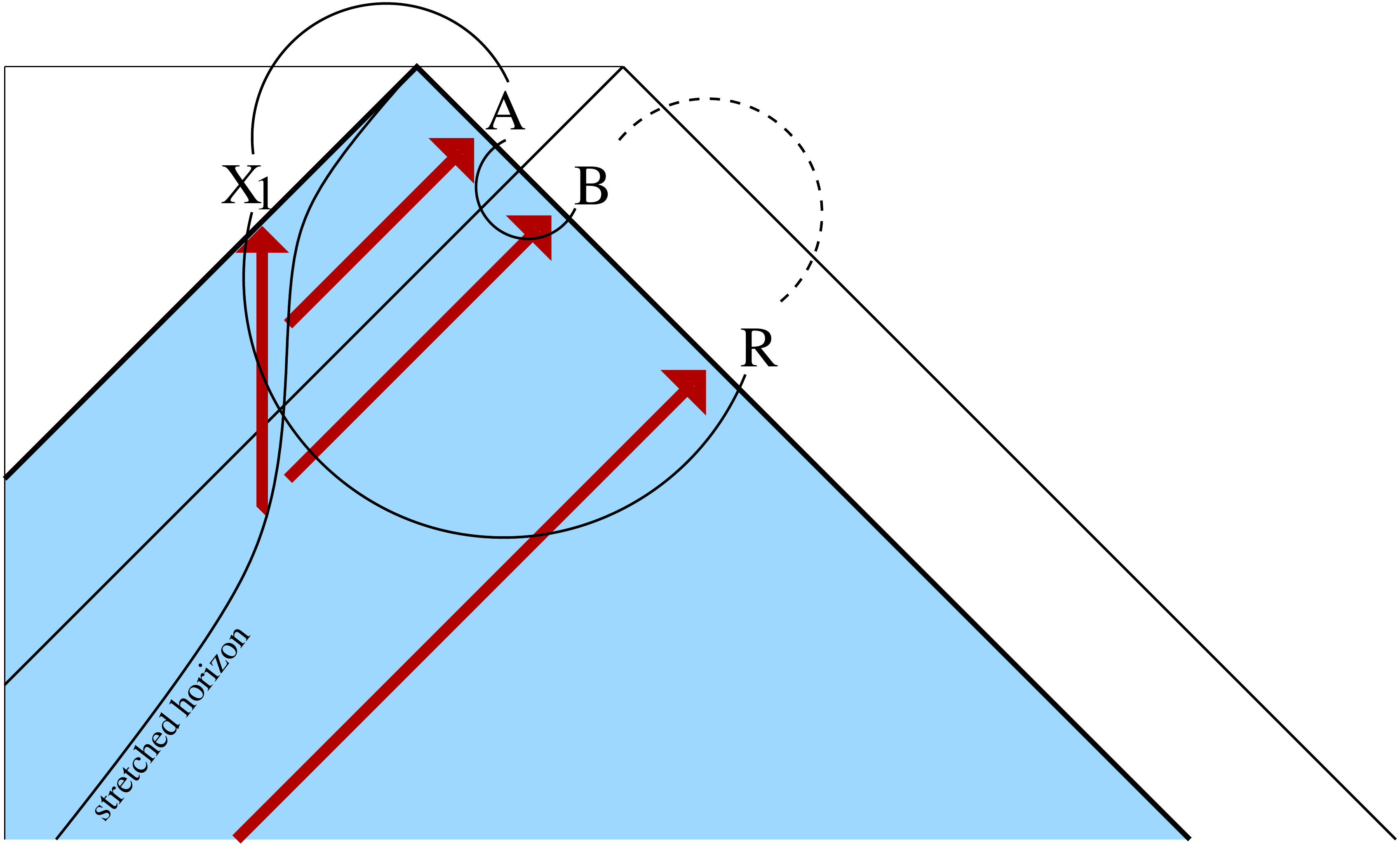}
\end{center}
\caption{An example of the Horowitz-Maldacena final-state proposal as applied to an observer (shaded causal patch) who enters the black hole at the Page time. Entanglement is denoted by semicircles. By the infall time, the initial state has evolves into an entangled state of the Hawking radiation $R$ and the half-evaporated black hole before $X_1$.  The initial state also imposes the infalling vacuum, so $A$ and $B$ are maximally entangled.  If the final state maximally entangles $X_1$ with $A$, then the quantum information in $X_1$ is teleported to $B$, and $BR$ will be in an entangled pure state.}
\label{fig-entswap}
\end{figure} 

In the context of firewalls, Kitaev and Preskill recently pointed out that the monogamy of entanglement can be violated in final-state quantum mechanics.\footnote{J.~Preskill, private communication.  The possibility that a variant of the HM proposal may play a role in resolving the firewall paradox was also raised in Ref.~\cite{AMPS}.}  Again, this can be understood in terms of quantum teleportation.  Consider four systems $X_1, X_2, X_3, X_4$, each with equal Hilbert space dimension, and let the initial state be a product of a maximally entangled state in ${\cal H}_{X_1}\otimes {\cal H}_{X_4}$ and a maximally entangled state in ${\cal H}_{X_2}\otimes {\cal H}_{X_3}$.  Let the final state maximally entangle ${\cal H}_{X_1}$ with ${\cal H}_{X_2}$ while imposing no restriction on $X_3$ and $X_4$.  Then $X_3$ will behave as if it is maximally entangled with $X_4$, i.e., they will be found to be correlated in any basis.  (This is known as ``entanglement swapping'' in quantum information theory.)  But if instead, $X_3$ and $X_4$ are measured in any basis, it is this pair that will be found to be correlated.  This would not be possible in standard quantum mechanics.

This property seems to be precisely what is needed, but there are some open questions. What subsystems of an old black hole and its Hawking radiation should be identified with $X_\nu$, $\nu=1,\ldots, 4$?  Figure~\ref{fig-entswap} shows a possibility that may be explored.  For simplicity, consider an observer Alice who falls into the black hole {\em at\/} the Page time, when ${\cal A} = {\cal A}_0/2$.  Before Alice crosses the horizon, we may be guided by the tenets of complementarity, as applied to an outside observer, and make no reference to the black hole interior.\footnote{D.~Harlow is also considering complementarity-motivated adaptations of the HM proposal to the firewall problem.}  The matter system $M$ evolves unitarily, first into a large black hole, and then, by partial evaporation, into a smaller black hole $X_1$ and the early Hawking radiation, $R\equiv X_4$.  These two systems are in a (nearly) maximally entangled pure state.  When Alice falls in, her perception of the vacuum state near the horizon requires that the interior fields $A\equiv X_2$ be maximally entangled with the near-horizon fields $B\equiv X_3$, in the pure Rindler state.  (The viewpoint taken here is that during free-fall, the outside viewpoint of the black hole as an object dissolves, and the modes $A$ and $B$ emerge as new degrees of freedom from the transplanckian regime, as in Hawking's calculation\cite{Haw75}.)  This characterizes the initial state of ${\cal H}_{X_1}\otimes {\cal H}_{X_2}\otimes {\cal H}_{X_3}\otimes {\cal H}_{X_4}$.  If the final state entangles $X_1$ with $X_2$, entanglement is transferred to the $X_3 X_4$ pair, which is $BR$ in this case.  Unitarity is recovered.  

More generally, one can consider an observer Alice who falls in at an arbitrary time. The unitarily evolved state of the matter system just before infall consists of the stretched horizon, $X_1$, and the Hawking radiation that has already been emitted, $X_4=R$. The systems $X_2=A$ and $X_3=B$ are the (sub-Planckian) quantum fields with support inside outside the horizon in Alice's causal patch.  As before, $\dim {\cal H}_{X_1}=\dim {\cal H}_{X_2}=\dim {\cal H}_{X_3}=\exp({\cal A}/4)$, where ${\cal A}$ is the area of the black hole at the infall time.  The Hilbert space dimension of the early radiation $R=X_4$ can be larger or smaller.  What is required is the maximal entanglement of the infalling vacuum, $X_2X_3$, and the purity of the Hawking radiation $X_3X_4$.  This is possible if the final state maximally entangles $X_1$ with $X_2$.

Another question concerns interactions.  If the final state is specified in terms of free fields, then interactions between the matter and the interior modes can spoil the HM proposal~\cite{GotPre03}.  The final state effectively becomes $\mbox{}_{M\otimes{\rm in}}\langle \Psi|U _{M\otimes{\rm in}}$, where the unitary $U_{M\otimes{\rm in}}$ encodes the effects of interactions among the two subsystems.  If $U_{M\otimes{\rm in}}$ entangles ${\cal H}_{\rm in}$ with ${\cal H}_M$, then it unentangles the effective final state, and the quantum teleportation of the matter state to the outside field is not faithful.  This can be compensated by including interactions in the definition of the final state.  Instead of Eq.~(\ref{eq-final}), define the final state to be
\begin{equation}
\mbox{}_{M\otimes{\rm in}}\langle \Psi|= \mbox{}_{M\otimes{\rm in}}\langle \Phi|  (S_M\otimes \mathbb{1}_{\rm in}) U_{M\otimes{\rm in}}^\dagger~,
\label{eq-final2}
\end{equation}
This amounts to entangling the matter and in-fields at the horizon.  More precisely, the final state undoes all interactions that take place among the two subsystems after horizon-crossing. 
 
In the complementarity-motivated application of the HM proposal discussed above, suppose that Alice falls in after the scrambling time.  Then the matter system that formed the black hole cannot interact directly with sub-Planckian interior modes in her causal patch.  However, in a more general setting, it is possible for additional matter (such as Alice, if she is a physical object) to enter the black hole at the infall time.  Such matter would be part of the Hilbert space $X_1$.  Thus, it will be necessary for the final state to be defined so as to undo all interactions between $X_1$ and $A$ (which by definition can only occur after horizon crossing).   In particular, any measurements of $A$ are undone by the final state.  This may also bear on a third question that must be addressed, namely whether final-state quantum mechanics, as applied to this problem, can lead to pathologies such as apparent violations of causality.\footnote{I am grateful to D.~Stanford for discussions of this issue.}

\acknowledgments I would like to thank B.~Freivogel, D.~Harlow, P.~Hayden, J.~Maldacena, D.~Marolf, J.~Polchinski, J.~Preskill, V.~Rosenhaus, D.~Stanford, L.~Susskind, and R.~Wald for many discussions, comments, and explanations. This work was supported by the Berkeley Center for Theoretical Physics, by the National Science Foundation (award numbers 0855653 and 0756174), by fqxi grant RFP3-1004, and by the U.S.\ Department of Energy under Contract DE-AC02-05CH11231.

\bibliographystyle{utcaps}
\bibliography{all}

\end{document}